\documentclass[twocolumn,superscriptaddress,showpacs,preprintnumbers,amsmath,amssymb,prb]{revtex4}
\usepackage[dvips]{graphicx}% Include figure files
\usepackage{dcolumn}% Align table columns on decimal point
\usepackage{bm}% bold math
\newcommand{\msr}{$\mu$SR}

\newcommand{\lvo}{LiV$_2$O$_4$}

\newcommand{\ysmn}{Y$_{1-x}$Sc$_x$Mn$_2$}

\begin{document}

\preprint{APS/123-QED}
\title{Quasi-one-dimensional spin dynamics in \lvo: one-to-three dimensional crossover as a possible origin of heavy fermion state}
% Force line breaks with \\

\author{Ryosuke Kadono}
\author{Akihiro Koda}
\affiliation{Muon Science Laboratory and Condensed Matter Research Center, Institute of Materials Structure Science, High Energy Accelerator Research Organization (KEK), Tsukuba, Ibaraki 305-0801, Japan}
\author{Wataru Higemoto}
\affiliation{Advanced Science Research Center, Japan Atomic Energy Agency, Tokai, Naka, Ibaraki 319-1195, Japan}
\author{Kazuki Ohishi}
\affiliation{Research Center for Neutron Science and Technology, Comprehensive Research Organization for Science and Society, Tokai, Naka, Ibaraki 319-1106, Japan} 
\author{\\ Hiroaki Ueda}
\affiliation{Department of Chemistry, Graduate School of Science, Kyoto University, Kyoto 605-8062, Japan}
\author{Chiharu Urano}
\thanks{Present address: National Institute of Advanced Industrial Science and Technology, Tsukuba 305-8561, Japan}
\author{Shin-ichiro Kondo}
\thanks{Present address: Materials Laboratories, SONY Corporation, Tokyo 108-0075, Japan}
\affiliation{Department of Advanced Materials Science, School of Frontier Sciences, University of Tokyo, Kashiwa, Chiba 277-8561, Japan}
\author{Minoru Nohara}
\affiliation{Department of Physics, Okayama University, Okayama 700-8530, Japan}
\author{Hidenori Takagi}
\affiliation{Department of Advanced Materials Science, School of Frontier Sciences, University of Tokyo, Kashiwa, Chiba 277-8561, Japan}


\begin{abstract}

Spin fluctuation in \lvo\ is revisited by examining the earlier result of muon spin rotation/relaxation measurements. Instead of a relationship for the localized electron limit, one for {\sl itinerant} electron systems between muon depolarization rate and spin fluctuation rate ($\nu_D$) is employed to re-analyze data, which reveals that $\nu_D$ varies linearly with temperature ($\nu_D\propto T$) over a range $10^8$--$10^{12}$ s$^{-1}$ for $0.02\le T<10^2$ K.  Such a linear-$T$ behavior as well as the magnitude of $\nu_D$  is fully consistent with that of the magnetic relaxation rate previously observed by inelastic neutron scattering (INS), demonstrating that \msr\ and INS have a common time window over the fluctuation spectrum. The linear-$T$ dependence of $\nu_D$ is understood as a specific feature predicted by a Hubbard model for intersecting one-dimensional (1D) chains.  This quasi-1D character, which is co-existent with enhanced uniform susceptibility at low temperatures, supports the scenario of 1D-to-3D crossover for the microscopic origin of heavy-fermion behavior in \lvo.

\end{abstract}

%Use showkeys class option if keyword
\maketitle

Heavy fermion (HF) behavior observed in a cubic vanadium spinel, \lvo, has been in the spotlight of broad interest,\cite{Kondo:97,Urano:00}  since it comprises one of remarkable examples in which only  $d$-orbital electrons are relevant to the phenomenon.  The formation of heavy quasiparticle (QP)  state below a characteristic temperature, $T^*\simeq20$ K, is suggested by large Sommerfeld coefficient ($\gamma\simeq 420$ mJ/molK$^2$) and other bulk properties that are strikingly similar to typical $f$-electron HF compounds.  Moreover, it has been inferred from the recent photoemission spectroscopy that a peak of the density of states (DOS)  just above the Fermi energy ($E_F$) develops for $T<T^*$, which may correspond to the QP peak typically found for the $f$-electron systems.\cite{Shimo:06}

While these observations seem to favor the Kondo mechanism established for the $f$-electron compounds as a common microscopic origin for the HF behavior in \lvo, approaches to support this scenario have been elusive.  Earlier theoretical attempt to project the $d$-electron states (1.5 per V$^{3.5+}$ ion) onto the Kondo model by splitting them into two sub-bands by electronic correlationhad to introduce unusually large Kondo coupling ($J_K\sim10^3$ K) to overcome competing effect of the Hund coupling.\cite{Anisimov:99,Kusunose:00}  Our \msr\ study on single-crystalline sample provided evidence against the formation of a spin-singlet state, as it showed presence of staggered vanadium moments at low temperatures far below $T^*$ that was interpreted as the Kondo temperature ($T_K$) in this scenario.\cite{Koda:04,Koda:05}  In the meantime, the importance of highly symmetric crystal structure and potential influence of geometrical frustration has been stressed by various authors, leading to a large variety of theoretical models.\cite{Lacroix:01,Fulde:01,Shannon:02,Burdin:02,Hopkinson:02,Fujimoto:02,Tsunetsugu:02,Yamashita:03,Laad:03,Arita:07}  

In the previous study using muon spin rotation and relaxation (\msr), we have shown on a powder specimen of \lvo\ that the observed \msr\ signal consists of two components characterized by different response of depolarization rate ($\lambda$) to external magnetic field ($H_0$).\cite{Koda:04}   In particular, the signal with  $\lambda$ showing least dependence on $H_0$ ($\lambda_D$, with a fractional yield $f\simeq0.4$) is mostly independent of temperature below $\sim10^2$ K, from which we suggested that the corresponding fluctuation rate derived from a general relation between $\lambda$ and $\nu$,
\begin{equation}
\lambda\simeq \frac{2\delta_\mu^2\nu}{\nu^2+\gamma_\mu^2 H_0^2},\label{rfd}
\end{equation}
is also independent of temperature  ($\nu_D>10^9$ s$^{-1}$).  In contrast, $\lambda$ associated with another signal ($\lambda_S$, with $1-f\simeq0.6$) is readily suppressed by $H_0$, which has been ascribed to slowly fluctuating local magnetic moments ($\nu_S\sim10^6$-$10^7$ s$^{-1}$).  Although the occurrence of such a phase separation has been confirmed by subsequent \msr\ study on high-quality single-crystalline samples,  the increased yield $f$ ($\simeq0.8$) strongly suggests that clarifying the origin of $\nu_D$ is essential to the understanding of electronic state in \lvo.\cite{Koda:05}

 Despite the fact that \lvo\ is metallic, the use of Eq.~(\ref{rfd}), which is valid in the limit of localized spins with $\nu$ determined by the local exchange interaction ($J\sim h\nu$), is presumed to be justified by the presence of staggered vanadium moments suggested by broad \msr\ linewidth at low temperatures (which is also in line with presumption of the Kondo scenario for $T>T^*$).\cite{Koda:04} Besides this, Eq.~(\ref{rfd}) may be regarded as a good approximation at high temperatures where the electronic state is subject to strong damping by phonos. However, our recent \msr\ study on another $d$-electron heavy fermion metal, \ysmn, has demonstrated that more precaution should be taken for the interpretation of depolarization at lower temperatures.\cite{Miyazaki:11}   

Here, we revisit the spin dynamics of \lvo\ on alternative basis that the staggered magnetic moments are carried by itinerant $d$ electrons.  The corresponding muon spin depolarization is described by a modified version of Eq.~(\ref{rfd}),
\begin{equation}
\lambda\simeq \frac{2\delta_\mu^2\nu}{\nu^2+\gamma_\mu^2 H_0^2}\cdot\frac{\chi k_BT}{N_A\mu_B^2},\label{rfdm}
\end{equation}
where the factors additional to Eq.~(\ref{rfd}) stems from electronic density of state at the Fermi level with $\chi$ being the magnetic susceptibility, $N_A$ the Avogadro number, and $\mu_B$ the Bohr magneton.\cite{Moriya:73,Hasegawa:74}  As a consequence, our reanalysis indicates that $\nu_D$ is linearly dependent on temperature ($\nu\propto T$) with frequency ranging from $10^8$--$10^{12}$ s$^{-1}$, which accidentally serves to cancel the $T$-dependence of $\lambda$.   The distilled behavior of $\nu_D$ turns out to be in excellent agreement with the previous result of inelastic neutron scattering (INS),\cite{Lee:01} strongly suggesting that both \msr\ and INS have been observing a common phenomenon.  Moreover, the linear-$T$ dependence of spin fluctuation, which is commonly observed in \ysmn, is understood as a property characteristic to  the spin-spin correlation of the intersecting Hubbard chains that simulate the pyrochlore lattice.\cite{Lee:03} This implies the crucial role of $t_{2g}$ orbitals as one-dimensional (1D) chains that are under a strong geometrical constraint of pyrochlore lattice structure, and further suggests the dimensional crossover due to coupling between these chains as the primary origin of the heavy-fermion state.\cite{Fujimoto:02}

\begin{figure}[t]
\begin{center}
\includegraphics[width=0.45\textwidth]{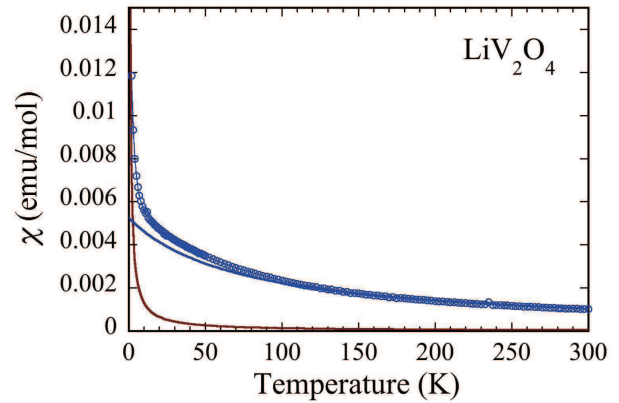}
\caption{(Color online)
Magnetic susceptibility ($\chi$) as a function of temperature in the powder specimen used for \msr\ measurement in Ref.\onlinecite{Koda:04}. Solid curve shows the best fit using a form described in the text.
}
\label{chi}
\end{center}
\end{figure}

Since the previous study on single-crystalline sample indicates that muon depolarization is dominated by a component showing fast fluctuation ($\nu_D$), we focus on the behavior of $\nu_D$.  The evaluation of $\nu_D$ using Eq.~(\ref{rfdm}) requires additional information on magnetic susceptibility ($\chi$) shown in Fig.~\ref{chi}, which has been obtained on the powder sample used for the previous \msr\ measurement.\cite{Koda:04}  Considering the behavior of $\chi$ observed in single-crystalline sample,\cite{Urano:00,Matsushita:05} we attribute the divergent behavior with $T\rightarrow 0$ to unknown paramagnetic impurities and decompose data into two parts,
\begin{eqnarray}
\chi & = & \chi_{\rm V} + \chi_{\rm imp} \nonumber\\
          & = & \frac{a}{T+\theta} + \frac{b}{T},\label{chit}
\end{eqnarray}
where $\theta$ is the Weiss temperature.  Curve fits using Eq.~(\ref{chit}) yields 
$a=0.387(5)$ emu$\cdot$K/mol, $\theta=74(2)$ K, and $b=0.0131(2)$ emu$\cdot$K/mol, in which the behavior of $\chi_{\rm V}$ is in good agreement with that of single crystals. Henceforth we use $\chi=\chi_{\rm V}$ for the input of Eq.~(\ref{rfdm}).   Assuming that the Curie term comes from free V$^{3.5+}$ spins (1.5$\mu_B$), the fractional yield of impurity phase estimated by $b/T$ is 1.6\% of the total volume. This is far smaller than that of primary \msr\ signals (either $f$ or $1-f$ with $f\simeq0.4$), indicating that the paramagnetism of the impurity phase is irrelevant to the interpretation of \msr\ data. 

Another important quantity in Eq.~(\ref{rfdm}) is the hyperfine parameter, $\delta_\mu$. The muon Knight shift measurements in both powder and single-crystalline samples yield $\delta_{\mu(D)}\simeq0.5\pm0.2$ GHz/$\mu_B$ that corresponds to the signal with $\nu_D$.\cite{Koda:04,Koda:05}  Apart from the large error due to broad linewidth, it is in good agreement with calculated value, $\delta_{\mu(D)}=0.143$ GHz/$\mu_B$, for muons that occupy a site in the center of cyclic vanadium hexamer (as inferred from the observed \msr\ linewidth due to {\sl nuclear} magnetic moments at high temperatures) and are subject to the magnetic dipolar fields from vanadium ions.\cite{Koda:04}  Since the experimental value of hyperfine parameter has large uncertainty due to broad linewidth, we use the calculated value for the evaluation of $\nu_D$ below.  

\begin{figure}[t]
\begin{center}
\includegraphics[width=0.4\textwidth]{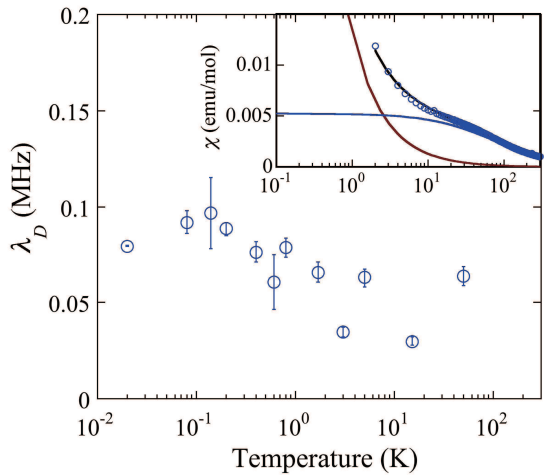}
\caption{(Color online)
Muon depolarization rate in \lvo\ reproduced using data published in Ref.\onlinecite{Koda:04} for the component showing fast fluctuation ($\nu_D$).
Inset: reproduction of Fig.~\ref{chi} with logarithmic scale for temperature.
}
\label{lmd}
\end{center}
\end{figure}

 Fig.~\ref{lmd} shows muon depolarization rate ($\lambda_D$) under zero external field, deduced by curve fits using a sum of two exponential damping,\cite{Koda:04} $$P_z(t)=f\exp(-\lambda_Dt)+(1-f)\exp(-\lambda_St).$$ 
 Although the data are scarce particularly at higher temperatures, $\lambda_D$ is 
only weakly dependent on temperature with a tendency of gradual increase with decreasing temperature, which is qualitatively similar to the behavior of $\chi_V$ (see the inset of Fig.~\ref{lmd}). These features are remarkably similar to that observed in \ysmn\ with greater Sc content ($x\ge0.07$).\cite{Miyazaki:11}

\begin{figure}[t]
\begin{center}
\includegraphics[width=0.45\textwidth]{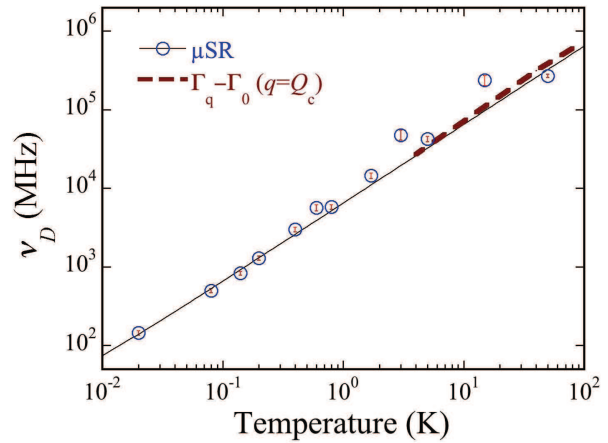}
\caption{(Color online)
Spin fluctuation rate ($\nu$) as a function of temperature in \lvo.  Thin solid line shows linear $T$ dependence ($\nu\propto T$).   Inelastic neutron scattering (INS) data are quoted for comparison, where the dashed curve shows the linewidth $(\Gamma_q-\Gamma_0)/h$ at $Q_c=0.64$ \AA$^{-1}$ with $\Gamma_0\simeq1.5$ meV.\cite{Lee:01} (See text for more detail.)
}
\label{nutemp}
\end{center}
\end{figure}

The evaluation of $\nu_D$ from $\lambda_D$ is straightforward by using Eq.~(\ref{rfdm}) in a modified form,
\begin{equation}
\nu_D\simeq \frac{2\delta_\mu^2}{\lambda_D}\cdot\frac{\chi k_BT}{N_A\mu_B^2},\label{nud}
\end{equation}
 where $H_0=0$ for the present condition of zero external field. 
 The result is plotted in Fig.~\ref{nutemp} together with data of INS,\cite{Lee:01}  where one can observe that $\nu_D$ falls on a straight line representing linear relation to temperature ($\nu_D\propto T$) over a $T$ range of three decades.  This is again strikingly similar to the behavior of spin fluctuation rate observed in \ysmn\ with $x\ge0.07$.\cite{Miyazaki:11}

 According to a numerical simulation of quarter-filled Hubbard chains (or rings) by path integral quantum Monte Carlo method, the spin dynamics is predicted to exhibit a linearly $T$-dependent relaxation rate in the spin-spin correlation,
\begin{equation}
\nu({\bf q}) = \nu_0 + c({\bf q})\cdot T,\label{nuq}
\end{equation} 
where $\nu_0$ comes from residual ferromagnetic spin correlation (corresponding wave vector ${\bf q}=0$).\cite{Lee:03} It is also inferred from the simulation that Eq.~(\ref{nuq}) depends only weakly on ${\bf q}$ and other parameters 
such as $U/t$ (where $U$ and $t$ are the on-site Coulomb energy and the transfer integral between nearest neighboring vanadium sites in the Hubbard model, respectively).  The observed behavior of $\nu_D$ is perfectly in line with the above prediction with $\nu_0\simeq0$.  As shown in Fig. 3, the magnitude of $\nu_D$  as well as its linear-$T$ dependence is in excellent agreement with the spin relaxation rate observed in INS at $q=Q_c=0.64$ \AA$^{-1}$ (around which a broad peak is observed below $T^*$) after subtracting the residual term, $\Gamma_q(T\rightarrow0)\equiv\Gamma_0\simeq1.5$ meV.\cite{Lee:01}  The difference between $\nu_0$ (\msr) and $\Gamma_0$ (INS)  may be attributed to that in the sensitive $q$ range of observation, as the $q$-integrated $\Gamma_q$ (for $0.6 <q<1.3$ \AA$^{-1}$) shows tendency of smaller $\Gamma_0$.\cite{Lee:01}  It must be remembered, however, that $\nu_D$ is subject to certain ambiguity due to the broadness of linewidth $\delta_{\mu(D)}$ observed in the muon Knight shift measurement.

While the comparison between $\nu_D$ and $\Gamma_q$ demonstrates that both \msr\ and INS probe nearly identical part of the spin fluctuation spectrum in \lvo, it also confirms that $^7$Li-NMR has a completely different time window of observation.\cite{Kondo:97,Johnston:05,Mahajan:98,Fujiwara:99,Trinkl:00,Kaps:01} Considering estimated hyperfine field on $^7$Li nuclei  ($A_{\rm hf}\simeq 0.258$ T$/\mu_B$ in Ref.\onlinecite{Mahajan:98}, or 0.171 T/$\mu_B$ in Ref.\onlinecite{Fujiwara:99}), the corresponding parameter $\delta_\mu$ in Eq.~(\ref{nud}) is evaluated to be $\gamma_{\rm Li}A_{\rm hf}\simeq17.8$-26.9 MHz/$\mu_B$ (where $\gamma_{\rm Li}=2\pi\times16.546$ MHz/T).  Since the spin-lattice relaxation rate ($1/T_1=\lambda$) is reported to vary over a range of $10^0$-$10^2$ s$^{-1}$ and to follow the Korringa relation ($\lambda/T\sim2.0$--2.25 s$^{-1}$K$^{-1}$) for $T<T^*\simeq20$ K,\cite{Kondo:97,Johnston:05,Mahajan:98,Fujiwara:99} the corresponding spin fluctuation rate evaluated by Eq.~(\ref{nud})  levels off to take a constant value, $\nu_{\rm Li}\sim10^{13}$ s$^{-1}$ ($T\ll T^*$). Interestingly, the estimation of $\nu_{\rm Li}$ using relationship between the electronic specific heat coefficient and $N(E_F)$, $\gamma=(2\pi^2/3)k_B^2N(E_F)N_A=0.420$ J/mol K$^2$ yields
\begin{equation}
\nu_{\rm Li}\simeq\frac{1}{\hbar N(E_F)} =\frac{2\pi^2k_B^2N_A}{3\hbar\gamma}=1.7\times10^{13}\:\:{\rm s}^{-1}.
\end{equation}
This is in excellent agreement with the experimental observation, and strongly suggests that $1/T_1$ in $^7$Li-NMR is dominated by the Pauli paramagnetism.
It should be stressed that the above frequency is beyond the limit of observation for \msr\ under the present condition, because the corresponding $\lambda$ would be less than 10$^{-2}$ MHz. 

A theoretical model by Fujimoto presumes the quasi-1D character of the $t_{2g}$ bands associated with the pyrochlore lattice (consisting of intersecting chains of $t_{2g}$ orbitals) as an essential basis for the description of electronic state in \lvo, because the hybridization between the 1D bands will be strongly suppressed owing to the geometrical configuration.\cite{Fujimoto:02}  It is interesting to note that such quasi-1D bands may have singularities in DOS, as in an analogous case of A15-type intermetallic compounds (e.g., V$_3$Si) where it is predicted that $N(E)\propto 1/(E-E_c)^{1/2}$ with $E_c$ being the position of the $d$ band.\cite{Labbe:67} The model incorporates the hybridization as a perturbation to the 1D Hubbard bands, which yields an energy scale ($T^*$) that characterizes the dimensional crossover from 1D to 3D as the Fermi-liquid state develops with decreasing temperature below $T^*$.  The calculated $\gamma$ taking account of the latter as the leading correction to the self-energy yields a large value, consistent with the experimentally observed ones.  The progression of hybridization also induces the enhancement of the 3D-like spin correlation that would appear as the enhancement of uniform susceptibility, while the spin fluctuation is dominated by the staggered component of 1D Hubbard chains.   The increase in $\chi$ ($=\chi_{\rm V}$) with decreasing temperature shown in Fig.~\ref{chi} is understood as the manifestation of such a dimensional crossover, as it perfectly follows the prediction by the relevant theory.\cite{Fujimoto:02}  The quasi-1D character of the low-energy spin fluctuation preserved  below $T^*$ coexisting with the enhanced $\chi$, which is commonly observed in \ysmn, comprises strong evidence of such a scenario.

It would be honest to mention that the muon depolarization rate ($\lambda$, from which $\nu$ is deduced) does not provide any criterion in itself to decide the credibility of the present interpretation based on itinerant model compared with that based on the local spin model adopted in our previous report.  However, the itinerant model has substantial merit as it provides a coherent understanding of spin fluctuation in metallic \lvo\ without resorting to the splitting of $t_{2g}$ band for a part of vanadium ions to bear ``local" spins, where such a band splitting is yet to be demonstrated experimentally.

Finally, we briefly mention the signal component with slower spin fluctuation observed by \msr.\cite{Koda:04,Koda:05}  This component is characterized by relatively small hyperfine parameter ($\delta_{\mu(S)}=23$-34 MHz/$\mu_B$), from which  fluctuation rate $\nu_S$ is evaluated by Eq.~(\ref{nud}) to yield $10^6$-$10^7$ s$^{-1}$ at low temperatures ($T<1$ K).  Considering the fact that the fractional yield of this component  decreases drastically in the single-crystalline samples ($1-f\simeq0.2$), one may be inclined to associate this component to some unknown impurity phase. However, there is no other experimental evidence for such a second phase having different chemical form with the relative yield as high as $\sim$20\%.  Thus more careful investigation would be needed to clarify the origin of this slow fluctuation.

%Here, we suggest a possibility of ``edge state" observed in the Haldane chain systems.  It is now established that \lvo\ is modeled as the quasi-1D Hubbard chains This quasistatic magnetism can be attributed to  staggered moments at the open edges of Haldane chains;\cite{Hagiwara:90} these edges may be due to various crystalline defects including the twin boundaries inherent in the orthorhombic structure.  A previous \msr\ study on another Haldane system (Y$_2$BaNiO$_5$, which has a similar gap energy) indicates that such staggered moments induce spin-glass-like random magnetism.\cite{Kojima:95}  

In conclusion, we have shown that reanalysis of muon depolarization rate in \lvo\ using the model of spin depolarization for the itinerant electron systems yields spin fluctuation rate that is linearly dependent on temperature. This result turns out to be fully consistent with that of the previous inelastic neutron scattering experiment, providing a coherent understanding of the spin dynamics by the theoretical model of intersecting 1D-Hubbard chains that simulates pyrochlore lattice.  The persistent quasi-1D spin dynamics coexists with the enhanced uniform susceptibility at lower temperature ($T<T^*$), which is common to another $d$-electron heavy-fermion system \ysmn.  These observations strongly suggest that geometrically constrained $t_{2g}$ band is the primary stage for the formation of heavy quasiparticles with  
the 1D-to-3D dimensional crossover as a possible mechanism of effective mass enhancement.

%We thank S. Fujimoto and H. Nakamura for helpful discussion.
 
%\begin{acknowledgments}
% We would like to thank the TRIUMF staff for their technical support during the $\mu$SR experiment. This work was partially supported by the KEK-MSL Inter-University Program for Oversea Muon Facilities and by a Grant-in-Aid for Creative Scientific Research on Priority Areas from the Ministry of Education, Culture, Sports, Science and Technology, Japan.
%\end{acknowledgments}

%\bibitem{Hagiwara:90} M. Hagiwara, K. Katsumata, I. Affleck, B. I. Halperin, and J. P. Renard,Phys. Rev. Lett. {\bf 65}, 3181 (1990).

%\bibitem{Kojima:95} K. Kojima, A. Keren, L. P. Le, G. M. Luke, B. Nachumi, W. D. Wu, Y. J. Uemura, K. Kiyono, S. Miyasaka, H. Takagi, and S. Uchida, Phys. Rev. Lett. {\bf 74}, 3471 (1995).

\end{document}